\documentstyle[12pt,psfig]{article}
\begin{document}
\begin{center}
\Large{\bf Single Photon Production in Relativistic Heavy Ion Collisions
and Quark Hadron Phase Transition\footnote
{Invited talk given at National Seminar on Nuclear Physics, July 26-29 1999,
 Institute of Physics, Bhubaneswar, India}
}
\vskip 0.2in

\large{Dinesh Kumar Srivastava }
\vskip 0.2in

\large{\em Variable Energy Cyclotron Centre,\\
 1/AF Bidhan Nagar, Calcutta 
700 064, India}

\vskip 0.2in

Abstract

\vskip 0.2in

\end{center}
We discuss the recent developments in the study of single photon
production in relativistic heavy ion collisions.
In particular their production
at SPS,  RHIC, and LHC energies is re-examined in view
of the results of Aurenche et al which show that
the rate of photon production from quark gluon plasma, evaluated
at the order of two loops far exceeds the rates evaluated at one-loop level
which have formed the basis of all the estimates of photons so far.
We find that the production of photons from quark matter could easily
out-shine those from the hadronic matter in certain ideal conditions.
We further show that the earlier results lending support to the
possibility of quark-hadron phase transition from the measured
yield of single photons in $S+Au$ collisions at CERN SPS remain valid
when an account is made for these developments.

\section{Introduction}

Single photons can be counted among the first signatures~\cite{first}
  which were proposed to verify the formation of
deconfined strongly interacting matter- namely the quark gluon plasma
(QGP). Along with dileptons- which will have similar origins, they
constitute electro-magnetic probes which are believed to reveal the
history of evolution of the plasma, through a (likely) mixed phase
and the hadronic phase, as they do not re-scatter once produced
and their production cross section is a strongly increasing 
function of temperature. During the QGP phase, the single photons
are believed to originate from Compton 
($q\,(\overline{q})\,g\,\rightarrow\,q\,(\overline{q})\,\gamma$)
and annihilation ($q\,\overline{q}\,\rightarrow\,g\,\gamma$)
processes~\cite{joe,rolf} as well as bremsstrahlung processes
($q\,q\,(g)\,\rightarrow\,q\,q\,(g)\,\gamma$). Recently in the
first evaluation of single photons within a parton cascade model~\cite{pcm},
it was shown~\cite{pcmphot} that the fragmentation of time-like 
quarks ($q\,\rightarrow\,q\,\gamma$) produced
in (semi)hard multiple scatterings during the pre-equilibrium phase 
of the collision  leads to a substantial production of photons (flash of
photons!), whose $p_T$ is decided by the $Q^2$ of the scatterings and not the
temperature, as in the above mentioned calculations.

The upper limit for production of single photons in $S+Au$ collisions 
at SPS energies~\cite{wa80} has been used by several authors to rule out simple
hadronic equations of states~\cite{prl} and the final results for the $Pb+Pb$
collisions at SPS energies are eagerly awaited. 

In a significant development Aurenche et al~\cite{pat} have recently
evaluated the production of photons in a QGP up to two loops and shown
that the bremsstrahlung process gives a contribution which is
similar in magnitude to the Compton and annihilation contributions evaluated 
up to the order of one loop earlier~\cite{joe,rolf}. This
is in contrast to the `expectations' that the bremsstrahlung contributions
drop rapidly with energy (see Ref.~\cite{kryz,dipali} for estimates
within a soft photon approximation). They also reported an entirely new
mechanism for the production of hard photons through the annihilation
of an off-mass shell quark and an anti-quark, where the off-mass shell
quark is a product of scattering with another quark or gluon and
which completely dominates the emission of hard photons. 

 In the following we first discuss the results of Srivastava~\cite{epjc1}
for $Pb+Pb$ collisions at SPS, RHIC, and LHC energies in view of the
recent findings of Aurenche et al. Next, we discuss the reanalysis of
the $S+Au$ data at SPS energy by Srivastava and Sinha~\cite{epjc2}.

 If confirmed, these results provide a very important confirmation
of the occurrence of quark hadron phase transition in
relativistic heavy ion collisions.

\begin{figure}[t]
\psfig{file=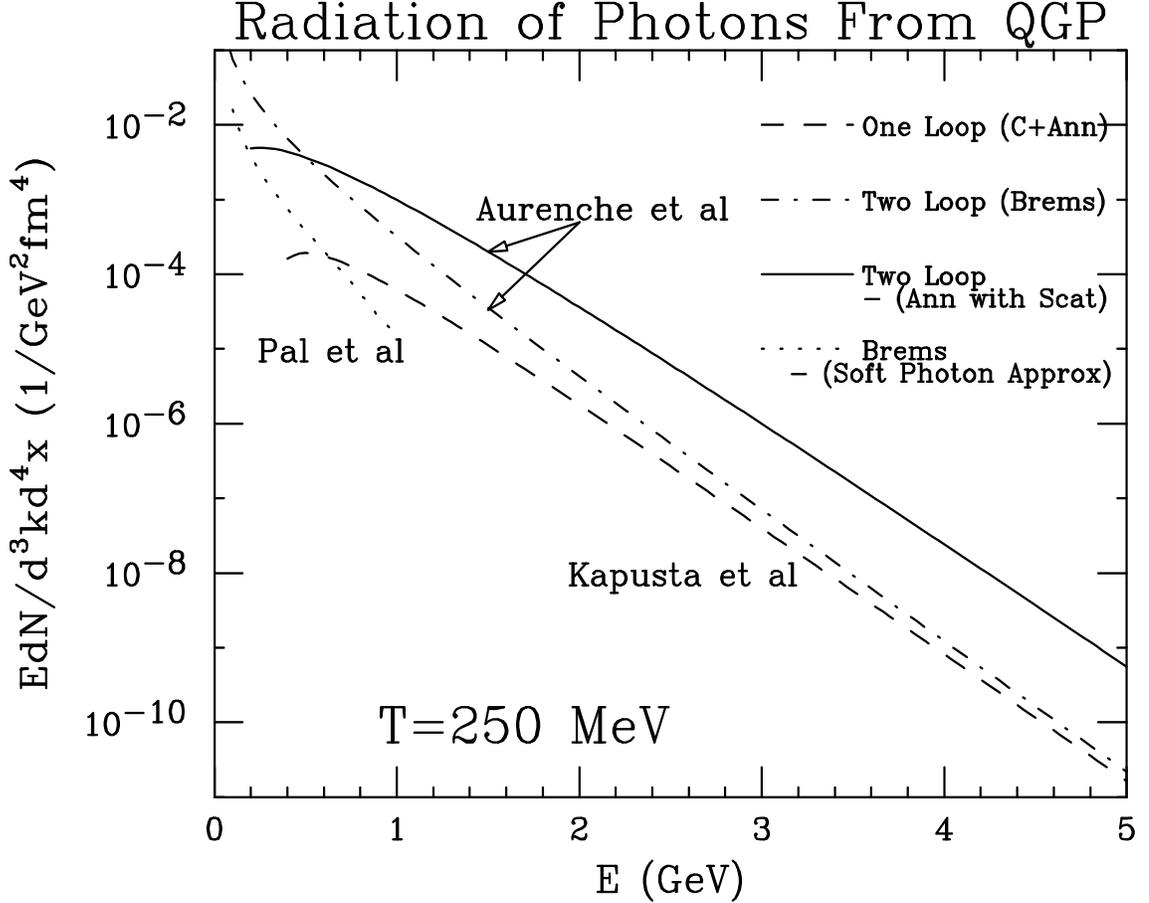,height=12cm,width=15cm}
\vskip 0.1in
\caption{ Radiation of photons from various processes in the quark
matter at $T=$ 250 MeV}
\end{figure}
\section{Results of Aurenche et al.}

Let us briefly recall the results of Aurenche et al and also earlier
work on single photon production from the quark matter.

The rate for the production of hard photons evaluated to one
loop order using the effective theory based on resummation of
hard thermal loops is given by~\cite{joe,rolf}:
\begin{equation}
E\frac{dN}{d^4x\,d^3k}=\frac{1}{2\pi^2}\,\alpha\alpha_s\,
                          \left(\sum_f e_f^2\right)\, T^2\,
                       e^{-E/T}\,\ln(\frac {cE}{\alpha_s T})
\end{equation}
where the constant $c\approx$ 0.23.  The summation runs over the
the flavours of the quarks and $e_f$ is the electric charge of the
quarks in units of charge of the electron. The rate of production
of photons due to the bremsstrahlung processes evaluated by
Aurenche et al is given by:
\begin{equation}
E\frac{dN}{d^4x\,d^3k}=\frac{8}{\pi^5}\,\alpha\alpha_s\,
                          \left(\sum_f e_f^2\right)\, 
                        \frac{T^4}{E^2}\,
                       e^{-E/T}\,(J_T-J_L)\,I(E,T)
\end{equation}
where $J_T\approx$ 4.45 and $J_L\approx - $4.26 for 2 flavours and 3
colour of quarks. For 3 flavour of quarks, $J_T\approx$ 4.80 and
 $J_L\approx - $4.52.  $I(E,T)$ stands for;
\begin{eqnarray}
I(E,T)&=&\left[ 3\zeta(3)+\frac{\pi^2}{6}\frac{E}{T}+
        \left(\frac{E}{T}\right)^2\ln(2)\right.\nonumber\\
        & &+4\,Li_3(-e^{-|E|/T})+2\,Li_2(-e^{-|E|/T})\nonumber\\
        & &\left. -\left(\frac{E}{T}\right)^2\,\ln(1+e^{-|E|/T})\right]~,
\end{eqnarray}
and the poly-logarith functions $Li$ are given by;
\begin{equation}
Li_a(z)=\sum_{n=1}^{+\infty}\frac{z^n}{n^a}~~.
\end{equation}

 And finally the contribution of the $q\overline{q}$ annihilation
with scattering obtained by them is given by:
\begin{equation}
E\frac{dN}{d^4x\,d^3k}=\frac{8}{3\pi^5}\,\alpha\alpha_s\,
                          \left(\sum_f e_f^2\right)\, ET \,
                          e^{-E/T}\,(J_T-J_L)
\end{equation}
 
We plot these rates of emission of photons
from a QGP at $T=$ 250 MeV (Fig.~1) for an easy comparison.
The dashed curve gives the contribution of the Compton
and annihilation processes evaluated to the order of one loop by
Kapusta et al~\cite{joe}, the dot-dashed curve gives the bremsstrahlung
contribution evaluated to two-loops by Aurenche et al~\cite{pat} while the
solid curve gives the results for the annihilation with scattering 
evaluated by the same authors. The dotted curve gives the results for
the bremsstrahlung contribution evaluated within a soft-photon
approximation (and using thermal mass for quarks and gluons) obtained
by Pal et al~\cite{dipali}. We see that at larger energies the
annihilation of quarks with scattering really dominates over the
rest of the contributions by more than a order of magnitude.

\section{$Pb+Pb$ collisions at SPS, RHIC, and LHC}

How much of this dominance does survive when we integrate the radiation of
photons over the history of evolution of the system, specially as the
QGP phase occurring in the early stages of the evolution necessarily
occupies smaller four-volume compared the hadronic matter,
which is known to have an emission rate similar to the quark matter
at a given temperature~\cite{joe} at least when only the Compton and the
annihilation terms are used?

In order to ascertain this we consider central collision of lead nuclei
at SPS, RHIC and LHC energies. We assume that a chemically and
thermally equilibrated quark-gluon plasma is formed at $\tau_0=$ 1
fm/$c$ at SPS and at 0.5 fm/$c$ at RHIC and LHC energies. While there are
indications that the plasma produced at the energies under
consideration may indeed attain thermal equilibrium at around
$\tau_0$ chosen here~\cite{pcm}, it is not quite definite
that it may be chemically equilibrated. It may be recalled 
that the parton cascade model which properly accounts for multiple
scatterings uses a cut-off in momentum transfer and virtuality to
regulate the divergences in the scattering  and the branching amplitudes for 
partons. This could underestimate the extent of chemical equilibration,
 by a cessation of interactions when the energy of the partons is still large
which would not be the case if the screening of the partonic 
interactions could be accounted for.
The self-screened parton cascade~\cite{sspc}
on the other hand attempts to remove these cut-offs by
estimating the screening offered by the partons which have larger
$p_T$ (and hence materialize earlier) to the partons which have
smaller $p_T$ (and hence materialize later). However it does not
explicitly account for multiple scattering except for what is contained 
in the Glauber approximation utilized there. 

In these exploratory calculations we assume a chemical equilibration
at the time $\tau_0$ such that the initial temperature is obtained
from the Bjorken condition~\cite{bj};
\begin{equation}
\frac{2\pi^4}{45\zeta(3)}\,\frac{1}{\pi R_T^2}\frac{dN}{dy}=4 a
T_0^3\tau_0
\end{equation}
where we have chosen the particle rapidity densities as 825, 1734,
and 5625 respectively at SPS, RHIC, and LHC energies for central
collision of lead nuclei~\cite{kms} and taken $a=47.5\pi^2/90$ for
a plasma of mass-less quarks (u, d, and s) and gluons.

We assume the phase transition to take place at $T=$ 160 MeV, and the
freeze-out to take place at 100 MeV.
We use a hadronic equation of state consisting of all the hadrons and
resonances from the particle data table which have a mass less then 2.5
GeV~\cite{jean}. The rates for the hadronic matter have been 
obtained~\cite{joe}
from a two loop approximation of the photon self energy 
using a model where $\pi-\rho$ interactions have been included. The 
contribution of the $A_1$ resonance is also included according to the
suggestions of Xiong et al~\cite{li}. The relevant hydrodynamic equations are
solved using the procedure~\cite{hydro} discussed earlier and
a integration over history of evolution is performed~\cite{jean}. 

In Fig.~2 we show our results for central collision of lead nuclei
at energies which are reached at CERN SPS. We give the
contribution of the quark
matter (from the QGP phase and the mixed phase) labeled as QM
and that of the hadronic matter (from the mixed phase and the hadronic
phase) separately.
 We see that if we use the rates obtained earlier by Kapusta et
al, there is no window when the radiations from the quark-matter could
shine above the contributions from the hadronic matter.  However, once the
newly obtained rates are used we see that the quark matter may 
indeed out-shine the hadronic matter up to $p_T=$ 2 GeV, from these
contributions alone.  
\begin{figure}
\psfig{file=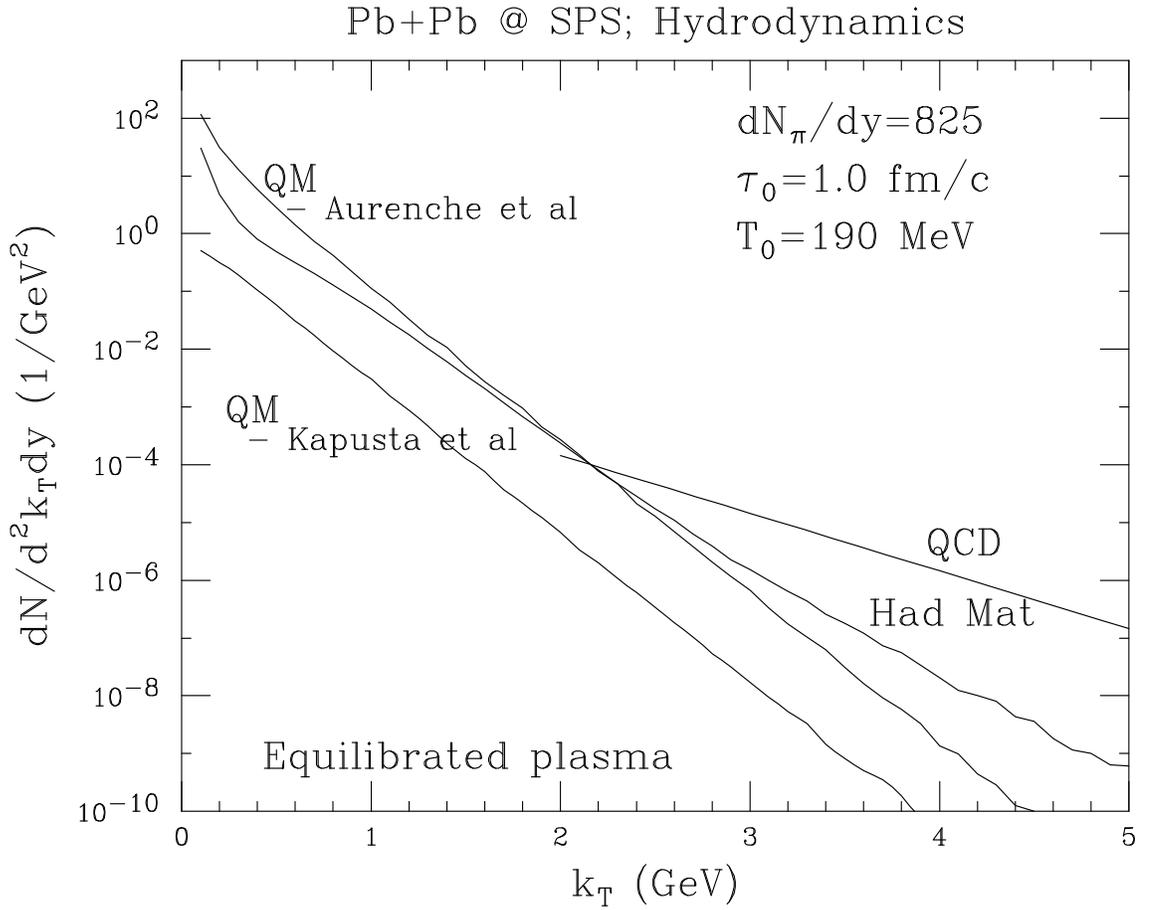,height=12cm,width=15cm}
\vskip 0.1in
\caption{ Radiation of photons from central collision of lead nuclei 
at SPS energies from the hadronic matter (in the mixed phase and the
hadronic phase) and the quark matter (in the QGP phase and the mixed
phase).
The contribution of the quark matter while using the
rates obtained by Kapusta et al and Aurenche et al,
and those from hard QCD processes 
 are shown separately
}
\end{figure}
\begin{figure}
\psfig{file=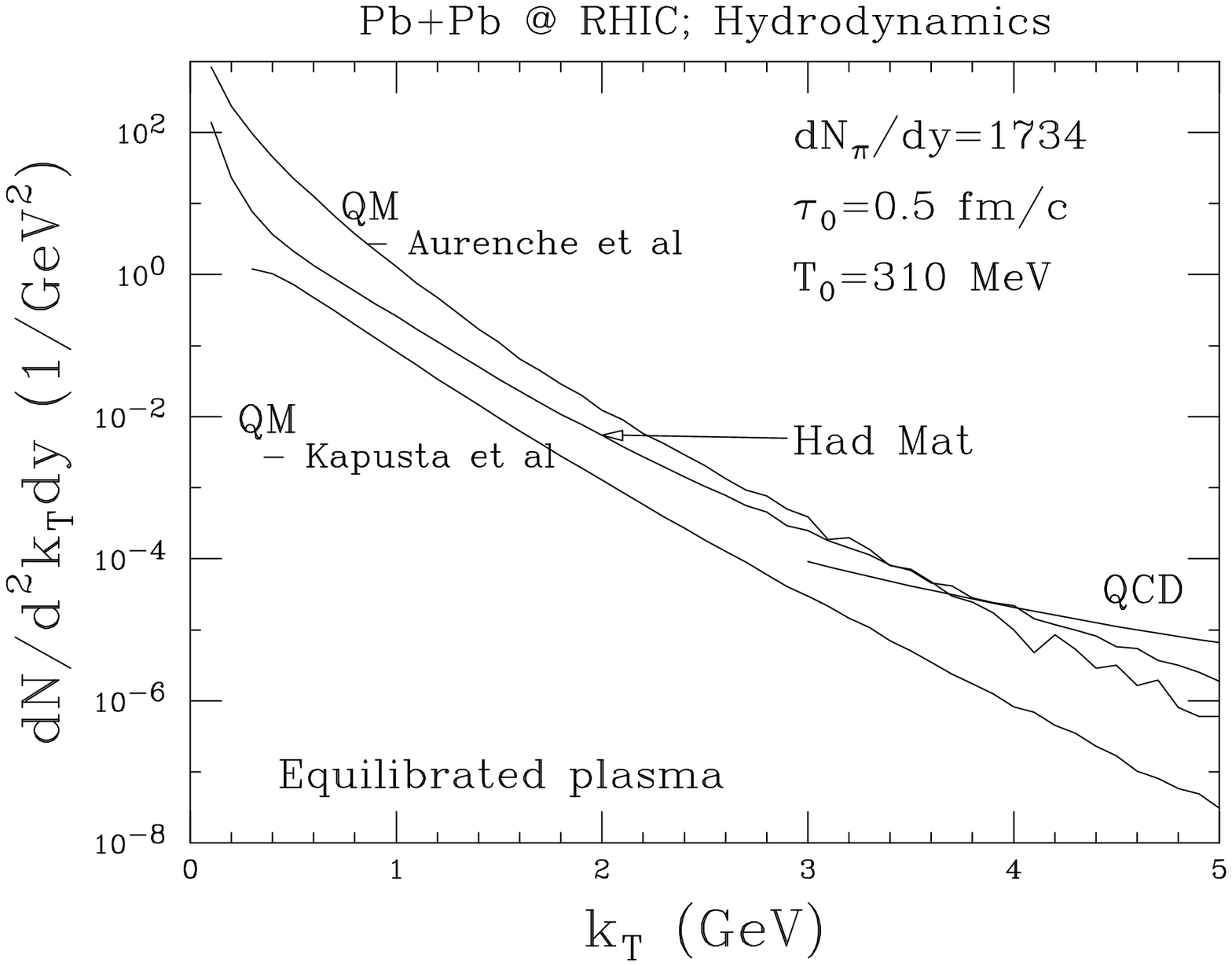,height=12cm,width=15cm}
\vskip 0.1in
\caption{ Same as Fig.~2 for RHIC energies.
}
\end{figure}
Note that by tracking the history from $\tau_0$= 1 fm/$c$ onward, we
have not included the pre-equilibrium contributions~\cite{pcmphot} which
will make a large contribution at higher momenta. 
The contribution of hard QCD photons~\cite{qcd} is obtained  by scaling 
the results for $pp$ collisions by the nuclear thickness.

The results for RHIC energies (Fig.~3) are quite interesting as now the window 
over which the quark matter out-shines the hadronic contributions
stretches to almost 3 GeV. Once again the addition of the pre-equilibrium
contributions at larger $p_T$ would substantially widen this window.

At LHC energies this window extends to beyond 4 GeV, and considering
that perhaps the local thermalization at LHC (and also at RHIC) could
be attained earlier than what is definitely a very conservative value
here, these results provide the exciting possibility that if these
conditions are met the quark matter may emit photons which may be
almost an order of magnitude larger than those coming from the
hadronic matter over a fairly wide window. As mentioned earlier, the
pre-equilibrium contribution (due to the very larger initial energy)
should be much larger here and we may have the exciting possibility that
the quark matter may out-shine the hadronic matter over a very large
window indeed.

\begin{figure}
\psfig{file=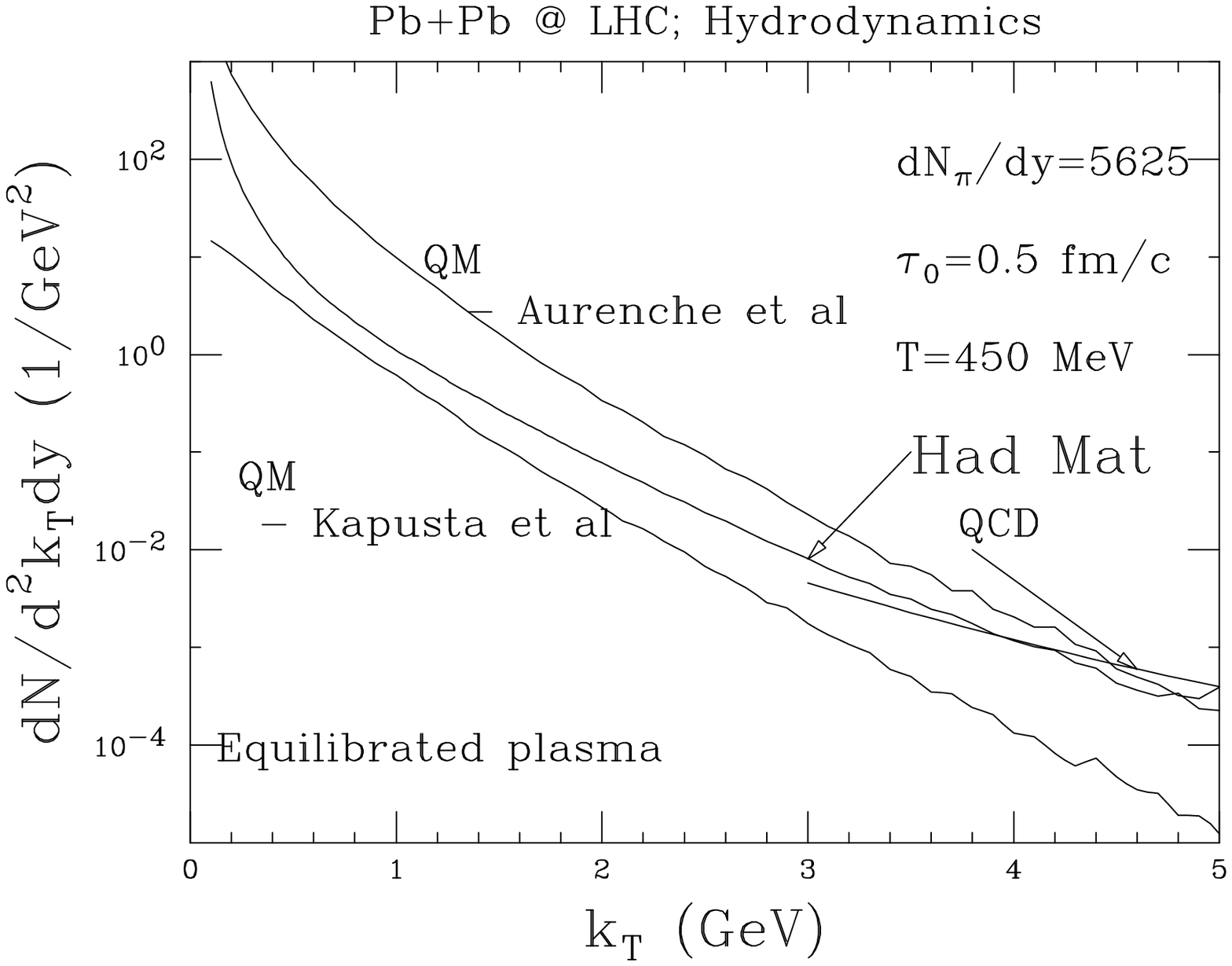,height=12cm,width=15cm}
\vskip 0.1in
\caption{ Same as Fig.~2 for LHC energies. 
}
\end{figure}

\section{Reanalysis of $S+Au$ collision at SPS}

The publication of the upper limit of the production of
single photons in $S+Au$ collisions at CERN SPS~\cite{wa80}
by the WA80 group has been preceded and followed by several papers
exploring their connection to the so-called quark-hadron phase
transition. Thus an early work by Srivastava and Sinha~\cite{prl},
 e.g., argued
that the data is consistent with a scenario where a quark-gluon
plasma is formed at some time $\tau_i\approx$ 1 fm/$c$, which expands
and cools, gets into a mixed phase of quarks, gluons, and hadrons,
and ultimately undergoes a freeze-out to a state of hadronic gas
consisting of $\pi$, $\rho$, $\omega$, and $\eta$ mesons. 
On the other hand, when the initial state is assumed to consist
of (the same) hadrons, the resulting large initial temperature leads to a
much larger production of single photons, in a gross
disagreement with the data.

\begin{figure}
\psfig{file=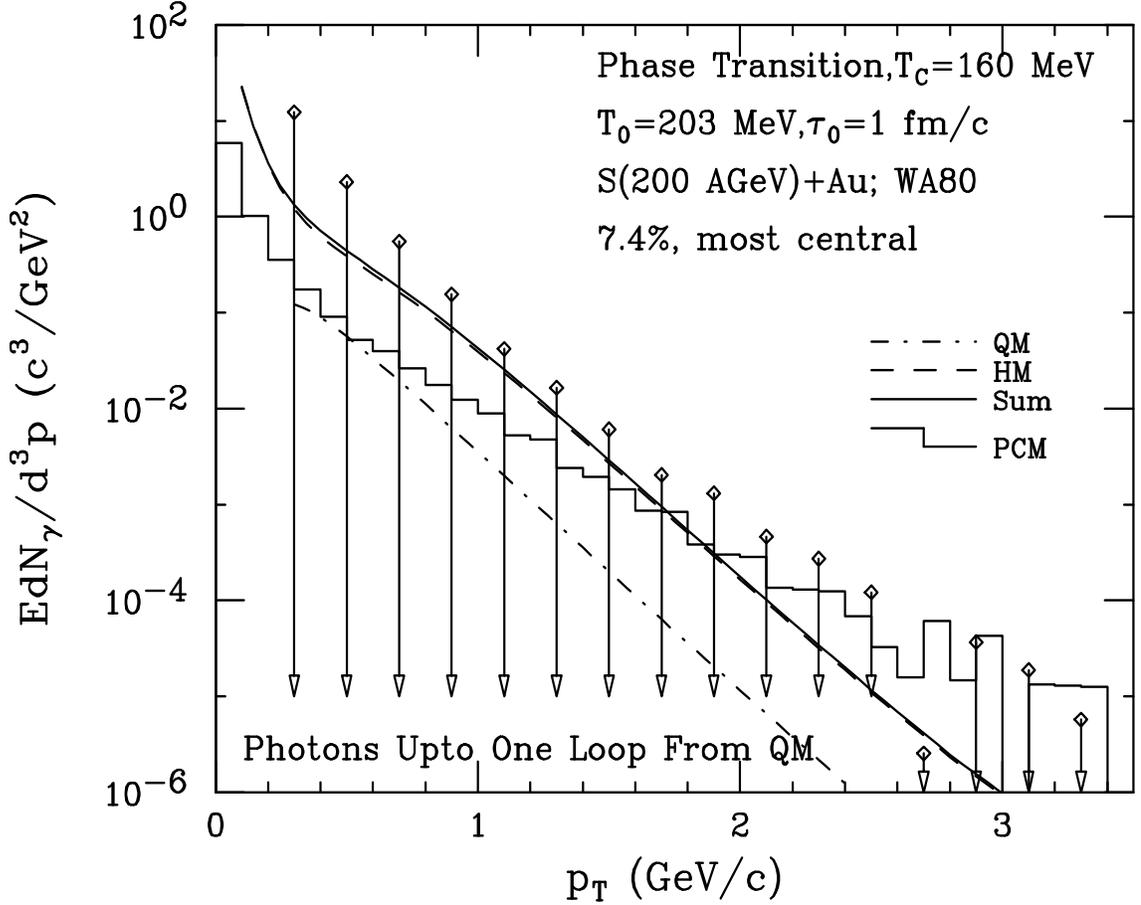,height=12cm,width=15cm}
\vskip 0.1in
\caption{ Single photon production in $S+Au$ collision at CERN SPS.
An equilibrated (chemically and thermally) quark-gluon plasma is
assumed to be formed at $\tau_0$ which expands, cools, gets into a
mixed phase and undergoes freeze-out. QM stands for radiations from the
quark matter in the QGP phase and the mixed phase. HM, likewise denotes
the radiation from the hadronic matter in the mixed phase and the
hadronic phase and Sum denotes the sum  of the contributions from the
equilibrium phase. The histogram shows the pre-equilibrium contribution
evaluated in a parton cascade model. The radiations from the 
quark-matter are evaluated to the order of one-loop.
}
\end{figure}

Since then, several authors have looked at the production of single
photons in such collisions, using varying evolution scenarios, and
including the effects of varying (baryon) density and temperature
on the rate of production of photons from the hadronic matter.

Thus for example, Cleymans, Redlich, and Srivastava~\cite{jean}
used a hadronic equation of state which included {\em {all}} hadrons having
a mass of upto 2.5 GeV, from the particle data book in complete
thermal and chemical equilibrium. In this approach, the production
of photons in phase-transition and no-phase transition scenarios
(for $Pb+Pb$ collisions at CERN SPS) was predicted to be quite 
similar in magnitude. However, the authors  also noted that, the no-phase
transition scenario necessitated a hadronic matter, where 2--3 hadrons
had to be accommodated within a volume of$\approx$ 1 fm$^3$, where the hadronic
picture should surely break-down.

All the above studies were performed using the (one-loop) evaluation
of single photons from the quark matter~\cite{joe,rolf} and hadronic reactions 
using varying effective Lagrangians. 

The findings of Aurenche et al provide a new dimension to these studies.
We have already seen that these findings provide that the
dominant number of the photons are now predicted to have their origin
in the quark matter if the initial state could be approximated
 as an an equilibrated plasma. 

\begin{figure}
\psfig{file=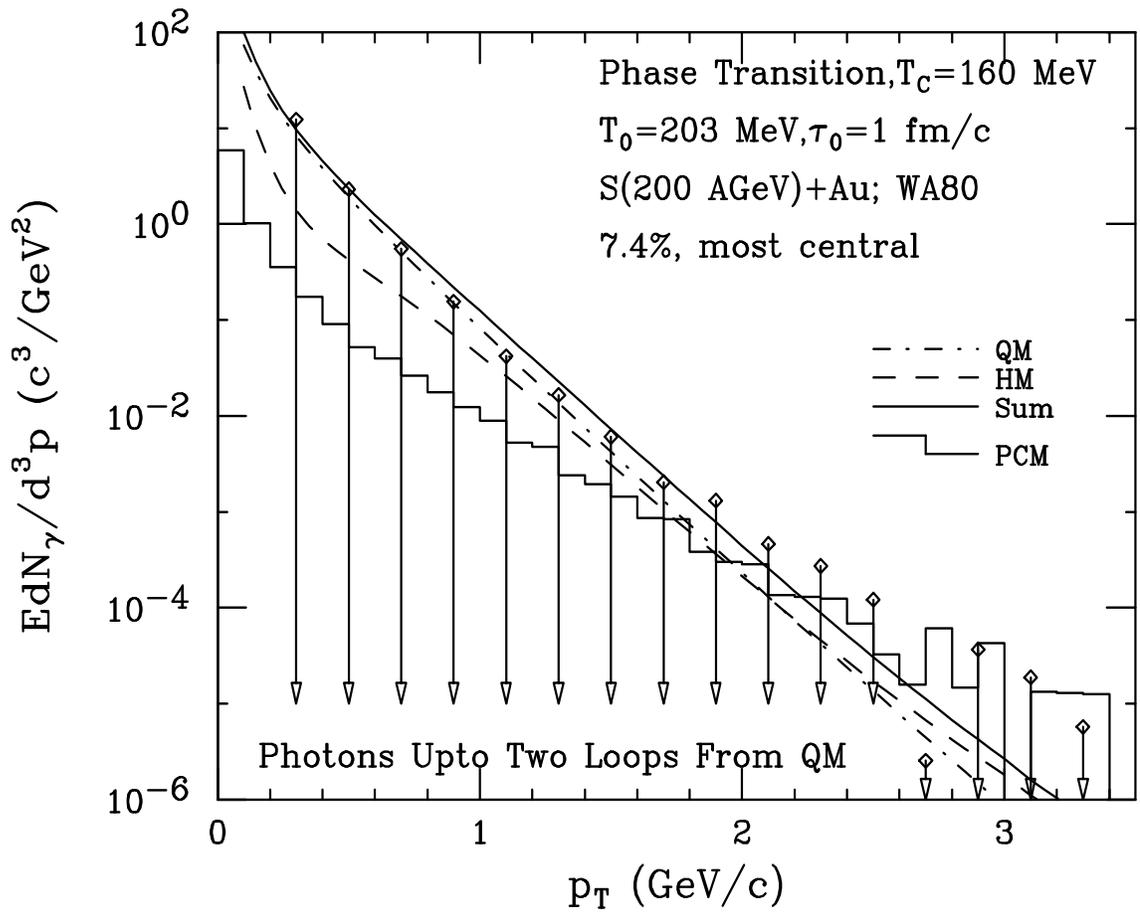,height=12cm,width=15cm}
\vskip 0.1in
\caption{ Same as Fig.~5, with the radiations from the
quark-matter evaluated to the order of two loops.
}
\end{figure}

This also raises a very important issue.  The 
analysis of Ref.~\cite{prl} has to be repeated to see if the
newly identified processes contributing to the single photons from
the quark matter remain consistent with the upper limit of the WA80
data.  We now discuss the outcome of this natural step, first
taken by Srivastava and Sinha~\cite{epjc2}.

These authors assumed, as in Ref.~\cite{prl}, that a chemically and thermally 
equilibrated quark-gluon
plasma was produced in the $S+Au$ collision at the time $\tau_0=$ 1 fm/$c$.
The Bjorken condition~\cite{bj} (Eq. 6) was then used to
get an estimate of the initial temperature. In the case of no 
phase transition the temperature  was obtained by demanding an yield of
the same entropy as when a QGP was assumed to be formed~\cite{jean}.
The rapidity density was taken as 225, 
and the transverse dimension was decided by the
radius of the $S$ nucleus.
Rest of the analysis followed along the lines discussed above.

In Fig.~5 we show  the  results  of Srivastava and Sinha~\cite{epjc2}
for the phase transition scenario.
As remarked in the figure caption there, the dot-dashed curve gives the
contribution of the quark-matter evaluated to the order of one loop,
 the dashed curve gives the
contribution of the hadronic matter, and the solid curve gives the
sum of the two. The 
{\em pre-equilibrium} contribution
evaluated within a parton cascade model~\cite{pcmphot} is also given.
 It is interesting  to see that the non-exponential component
apparent in the measured upper limit can be identified with the
 pre-equilibrium contribution.

It is seen that the photon yield stays below the upper limit at all
$p_T$, and most significantly, the dominant contribution is
from the radiation from the hadronic matter.

The corresponding results with rates evaluated to the order of two-loops
are given in Fig.~6 using similar notations.
We now see that the evaluated photon yield has a dominant
contribution from the quark-matter, as remarked earlier~\cite{epjc1}.
 We also note that the predicted yield closely follows the shape of the
measured upper limit over the entire range of $p_T$.

Note that the evaluated photon yield exhausts the upper limit at all
$p_T$. However, considering that the measurements represent an upper limit,
it still leaves a scope for a discussion of scenarios which may reduce the
yield of single photons. The fore-most consideration, and which
is also most likely, would be an initial state where
 the quark-gluon plasma is {\em not} in chemical
equilibrium.

In Fig.~7 we have shown the predictions for the scenario when
no phase transition takes place.  We again see that at least beyond
$p_T$ equal to 1 GeV/$c$, the  estimated single photon yield 
is consistent with the upper limit, though it is smaller than the
upper limit both at the lower and the upper end of the $p_T$
spectrum. However, we have to emphasize that this description
involves a hadronic gas which has a number density of several
hadrons/fm$^3$, which is rather un-physical, and 
we have reservations about this 
description.
 We may also add that for such high
hadronic densities almost all prescriptions for accommodating
finite size effects in the hadronic equation of state will either
break-down or imply a very high energy density to overcome the
so-called hard-core repulsion of the hadrons at very short
densities.  We can thus be fairly confident that the picture leading to
the Fig.~6 (or 5) is more likely.

Before concluding, it is of interest to add a comment on the shape of the
predicted spectra (Figs.~5--7) in comparison to the measured
upper limit. We have already
remarked that the predictions lie considerably below the upper
limit in Figs. 5 and
7, at lower $p_T$. Recall that the hadronic reactions considered in the
study include the process $\pi \pi \rightarrow \rho \gamma$ which
is known to be equivalent to the bremsstrahlung process $\pi \pi \rightarrow
\pi \pi \gamma$ (see Ref.~\cite{kryz,dipali}). Thus we realize that
the bremsstrahlung process in the quark matter contained in the two-loop
evaluations of Aurenche et al~\cite{pat} plays an important role
in getting the right shape of the spectrum at lower $p_T$. We do not have
to repeat that the pre-equilibrium contribution leads to the right shape
at higher $p_T$.  

It is thus clear that the newly obtained rates for emission of
photons from QGP (evaluated to the order of two loops), which are
much larger than the corresponding results for the one-loop
estimates yield single photons which are in agreement with the
upper limit of the data obtained by the WA80 experiment for the
$S+Au$ collisions at CERN SPS, and support a description
where a quark-gluon plasma is formed. 

We repeat also that, considering that the data represent the
upper limit, we can, in principle, admit a scenario  which has a chemically
non-equilibrated plasma at the time $\tau_0$, and which
will lead to a smaller radiation of photons from the quark phase.

\begin{figure}
\psfig{file=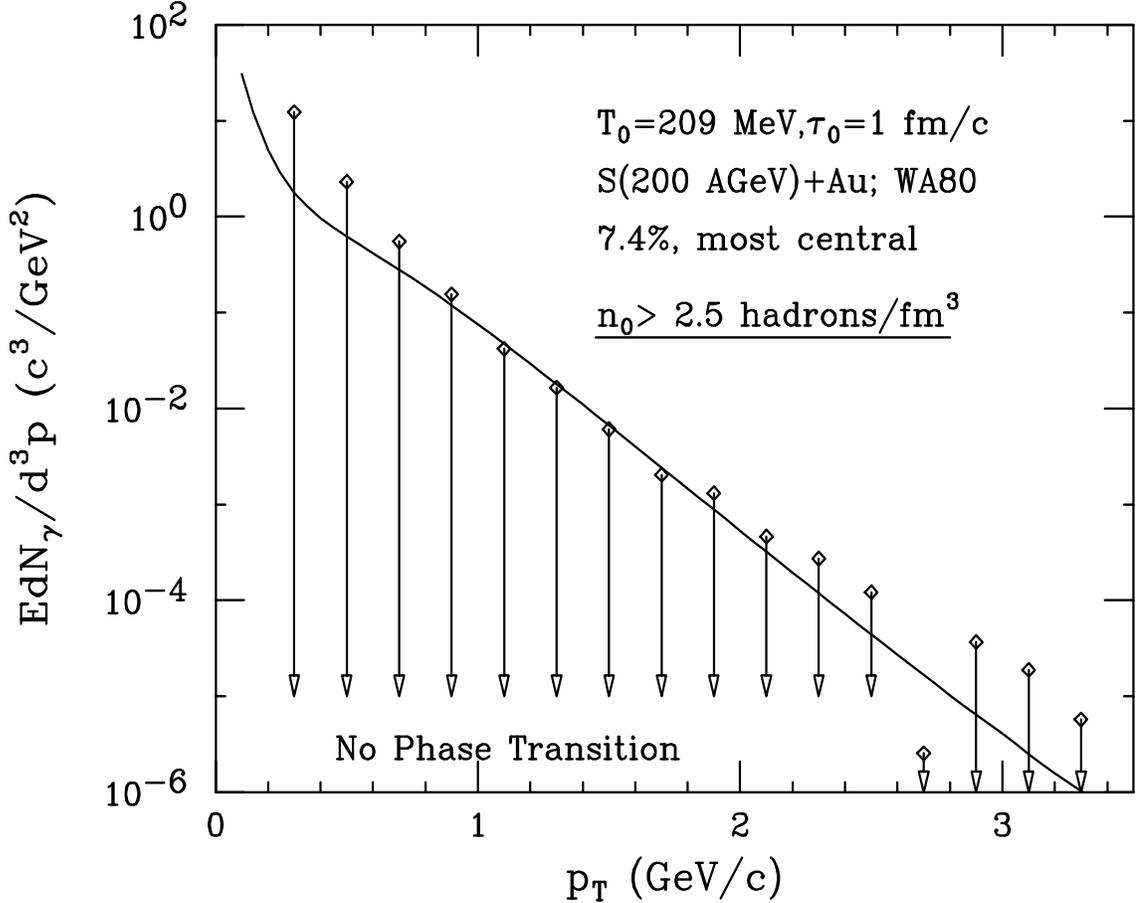,height=12cm,width=15cm}
\vskip 0.1in
\caption{ Same as Fig.~5, but without a phase transition; i.e.,
a hot hadronic gas is assumed to be formed at $\tau_0$. Note, however,
in this case the initial density of hadrons exceeds several
hadrons/fm$^3$.
}
\end{figure}
\section{Discussions and Summary}

How will the results change if the QGP is not in chemical equilibrium?
While it is not easy to perform the estimates similar to the one 
done by Aurenche et al for a chemically non-equilibrated plasma,
it is reasonable to assume that the rates will fall simply because
then the number of quarks and gluons will be smaller. Some of this
short-fall will be off-set by the much larger temperatures 
which the parton cascade models predict. If one considers
a chemically equilibrating plasma~\cite{smm} then the quark and
gluon fugacities will increase with time and at least the
contributions from the latter stages will not be strongly suppressed.
It is still felt that the loss of production of high $p_T$ (from
early times) photons due to chemical non-equilibration would be
more than off-set by the increased temperature  and the pre-equilibrium
contribution, which can be quite large.  

We conclude that the newly obtained rates for emission of
photons from QGP (evaluated to the order of two loops) suggest that
if chemically equilibrated plasma is produced then there will exist
a fairly wide window where the photons from quark matter
may outshine the photons from hadronic matter. Even in the absence
of chemical equilibration these results indicate an enhanced
radiation from the quark matter which is of considerable interest.

We have also seen that these development continue to support the
exciting possibility that the quark-hadron phase transition may have
indeed taken place in $S+Au$ collisions at SPS. The preliminary data
for $Pb+Pb$ collisions at SPS also seem to indicate~\cite{pcmphot} 
a similar possibility. This coupled with the observed excess production of
dileptons, excess production of strange particles and the suppression
of $J/\psi$ production, all of which constitute  signatures of the
quark hadron phase transition indicate that we are on the verge
of very clear confirmation of the discovery of quark gluon plasma.
A strong support to these observations has recently been provided
by the first estimate of the colour Debye mass~\cite{debye} for the partons
released in heavy ion collisions evaluated within the parton
cascade model. For $Pb+Pb$ collisions at SPS energies, it is found to be larger
than the screening mass of $\chi_c$ signalling its dissolution and the
attendant reduction in the production of $J/\psi$. For higher energies,
the Debye mass is found to be much larger.

\section*{Acknowledgments} 
The author gratefully acknowledges the encouragement and thoughtful
guidance he has received from Prof. L. Satpathy and Prof. M. Satpathy
 over the last two decades. The work described here has been done in 
collaboration with Bikash Sinha.

\end{document}